\documentclass[conference]{IEEEtran}
\IEEEoverridecommandlockouts

\usepackage{cite}
\usepackage{amsmath,amssymb,amsfonts}
\usepackage{algorithm}
\usepackage{algorithmic}
\usepackage{graphicx}
\usepackage{textcomp}
\usepackage{xcolor}
\usepackage{caption}
\usepackage{subcaption}
\usepackage{pgfplots}
\usepackage{tikz}
\usepgfplotslibrary{groupplots}

\DeclareMathOperator{\RE}{\mathbb{R}} 

\def\BibTeX{{\rm B\kern-.05em{\sc i\kern-.025em b}\kern-.08em
    T\kern-.1667em\lower.7ex\hbox{E}\kern-.125emX}}

\begin{document}

\title{Power Failure Cascade Prediction using Graph Neural Networks}
\author{
	\IEEEauthorblockN{Sathwik Chadaga, Xinyu Wu, and Eytan Modiano}
	\IEEEauthorblockA{Laboratory for Information and Decision Systems, Massachusetts Institute of Technology, Cambridge, MA}
}

\maketitle

\begin{abstract}
	We consider the problem of predicting power failure cascades due to branch failures. We propose a flow-free model based on graph neural networks that predicts grid states at every generation of a cascade process given an initial contingency and power injection values. We train the proposed model using a cascade sequence data pool generated from simulations. We then evaluate our model at various levels of granularity. We present several error metrics that gauge the model's ability to predict the failure size, the final grid state, and the failure time steps of each branch within the cascade. We benchmark the graph neural network model against influence models. We show that, in addition to being generic over randomly scaled power injection values, the graph neural network model outperforms multiple influence models that are built specifically for their corresponding loading profiles. Finally, we show that the proposed model reduces the computational time by almost two orders of magnitude.
\end{abstract}


\section{Introduction}
Modern power grids often experience unpredictable component failures that are caused due to an exogenous event like a tree branch falling, bad weather, failure of an aged device, or an operator error. These random failures, if not treated properly, can propagate rapidly through the grid, potentially resulting in large scale blackouts. Hence, it is important to study such \emph{failure cascades} as part of the power contingency analysis. Moreover, power grids have seen a recent surge in outages \cite{increasedoutages} due to extreme weather conditions \cite{weatherNews} and power grid aging \cite{aginggrid}, causing significant losses to businesses, industries, and healthcare sectors  \cite{outagecost, outagecosthealth}, making the study of power failures increasingly important. 

There have been several studies performed on historical failure cascade data \cite{flowFreeInfluenceHines, flowFreeCascadeBranching2}. However, the scarce historical records of cascading failures are not representative of all the possibilities. Hence, numerical simulations and analysis methods have been proposed, which solve the static power flow problem step-by-step and determine the sequence of quasi-static transmission link overflows \cite{ pfBasedCascade2}. However, the AC power flow model is computationally expensive, while the computationally tractable DC power flow model has been shown to underestimate the failure sizes \cite{pfBasedCascadeACDC}.


To overcome the high complexity of flow-based methods, efforts have been devoted to constructing flow-free models of failure cascades. Tools like branching process  \cite{flowFreeCascadeBranching2}, the random chemistry algorithm \cite{flowFreeRandomChem1}, and the expectation-maximization algorithm \cite{flowFreeEM}  have been proposed to estimate blackout risks. These flow-free models aspire to capture the cascade flow dynamics from data, obtained either from simulations or historic outage records. This data driven approach has led researchers to investigate fast and accurate machine learning models.


Machine learning has been used in power system analysis in various settings \cite{MLinPFCSurvey1}. For example, as power flow calculation using Newton-Raphson is computationally expensive, more efficient power flow calculation methods have been proposed using deep \cite{dnnPowerFlow} and convolutional  \cite{cnnPowerFlow} neural networks. Moreover, in the area of cascade prediction, support vector machines have been employed in blackout prediction \cite{blackoutPredSVM}, cascade failure size estimation \cite{cascadeSizePredSVMKNNetc}, and load loss estimation \cite{loadLossFracPredSVM}. 
Additionally, methods using Bayes networks have been proposed for failure cascade prediction in \cite{cascadeSeqPredBayesNet}. Despite being computationally efficient, these techniques fail to take advantage of the power grid topology information leading us to explore techniques that use graph neural networks.

Graph neural networks (GNNs) are a type of neural networks that operate on graph-structured data \cite{ kipfGNN}. They process input graphs by repetitively updating the information at each node based on its neighbors, thereby leveraging the underlying graph topology. There have been recent applications of GNNs in the field of power networks. One such application is the design of computationally efficient power flow solvers. In \cite{supervisedPFGNN1, supervisedPFGNN2, supervisedPFGNN3}, GNNs are trained in a supervised way to imitate the Newton-Raphson power flow solver. Whereas \cite{unsupervisedPFGNN1, unsupervisedPFGNN2} follow an unsupervised learning method that minimizes the violation of Kirchoff's laws. 

Moreover, GNNs have seen recent applications in the field of power failure cascades. GNNs have been used for real time grid monitoring tasks during a cascade, like predictions of optimal load shedding \cite{realTimeOptimalLoadShedding} and total load lost \cite{realTimeLoadLossEstimation}. 
These works involve a graph-level prediction task, i.e. they predict a particular property of the grid as a whole. GNNs can also be used for edge-level and node-level prediction tasks. For example, in \cite{avalancheCentrality}, a node-level vulnerability metric called the Avalanche Centrality is predicted for all nodes of the grid using GNNs. In \cite{FCSearch2}, efficient failure cascade path search techniques with GNNs have been proposed.

The existing works as discussed above are focused on characterizing one or two aspects of failure cascades, like load loss, failure size, or blackout possibility, lacking a comprehensive evaluation of the cascade at finer levels of granularity. This is addressed in \cite{ flowFreeInfluenceXinyu}, where an influence model is trained to predict the power grid states within a cascade. However, the influence model approach cannot generalize for variable loading as it does not take as input the power injection values. A flow-based GNN model has been proposed in \cite{physicsGNNforCascadePred} that can generalize for variable power injections. However, this model is centered around predicting the power flow values in a step-by-step manner to obtain the sequence of branch overflows. Hence, even though this technique speeds up the cascade prediction process compared to traditional methods, it still involves a high computational overhead in handling the formation of islands during the cascade, such as identification of islands and rebalancing the load and power generation within islands.

In this paper, we build a flow-free GNN model that does cascade sequence prediction without requiring power flow calculation at every generation of the cascade. We summarize our contributions below.
\begin{enumerate}
	\item We propose a flow-free model based on a GNN that predicts grid states at every generation of a cascade,  providing a way to comprehensively evaluate cascades at various levels of granularity. The proposed model takes as input the node power injection values, the initial contingency, and the grid topology. We use the cascading failure simulator oracle from \cite{flowFreeRandomChem1} to generate a cascade sequence dataset to train our model.
	\item We evaluate the performance of our model at various levels of granularity including prediction of the failure size, the final grid state, and the generations at which each branch fails within a cascade. We benchmark our model against the influence model  \cite{flowFreeInfluenceXinyu} and show that in addition to being generic over randomly scaled loading values, the GNN model outperforms different load-specific influence models under every metric.
	\item We perform a runtime analysis and show that the GNN model reduces the prediction time by almost two orders of magnitude compared to the DC power flow calculation based simulators.
\end{enumerate}

The rest of the paper is organized as follows. We formulate the problem of failure cascade prediction and describe the proposed graph neural network model in Section \ref{sec:gnn}. We discuss the cascading failure simulator oracle in Section \ref{sec:datasynth}. We present the model performance results in Section \ref{sec:results}.

\section{Problem Formulation and the GNN Model}\label{sec:gnn}
We consider the power failure cascade process due to branch failures. In this setting, a failure cascade begins with an initial failure of one or more branches in the grid. The initial branch failures perturb the power flow in the grid, leading other branches to overload and trip. The new failures further cause additional branches to trip and so on, consequently triggering a cascade process. The cascade process can be grouped into generations in time \cite{flowFreeCascadeBranching2}, which we refer to as time steps.

We represent the power grid by a directed graph ${G} = (V, E)$, where the nodes ${V}$ represent buses and the directed edges ${E}$ represent branches (or edges). For a branch $e \in E$ at time $t$, we choose the \emph{branch state} $s_e[t]$ to be its binary operational state, which can either be 0 (failed) or 1 (active). We define the \emph{network state} at time $t$ as $s[t] := (s_e[t])_{e\in E}$. 

Given the \emph{initial contingency} $s[0]$ (the network state at $t=0$), our goal is to predict the  \emph{cascade sequence} $s:=(s[t])_{t=0}^{T-1}$, where $T$ is the cascade length. However, we assume that once a branch fails, it stays in the failed state for the rest of the cascade. This allows us to define the \emph{failure step} of a branch $e$, the time step at which its state changes from 1 to 0, as $f_e := \sum_{t=0}^{T-1}s_e[t]$. From this failure step $f_e$, we can fully recover the branch states $s_e[t]$, and hence $s$, by setting $s_e[t] = 1$ for $0 \leq t<f_e$ and $s_e[t]=0$ for $f_e \leq t < T$ for all $e\in E$.  Hence, instead of predicting $s$ directly, we design a model that predicts the branch failure steps $f := (f_e)_{e\in E}$. We propose a GNN model to do this as explained below.

\begin{figure*}[tbp]
	\centering
	\includegraphics[width=\textwidth]{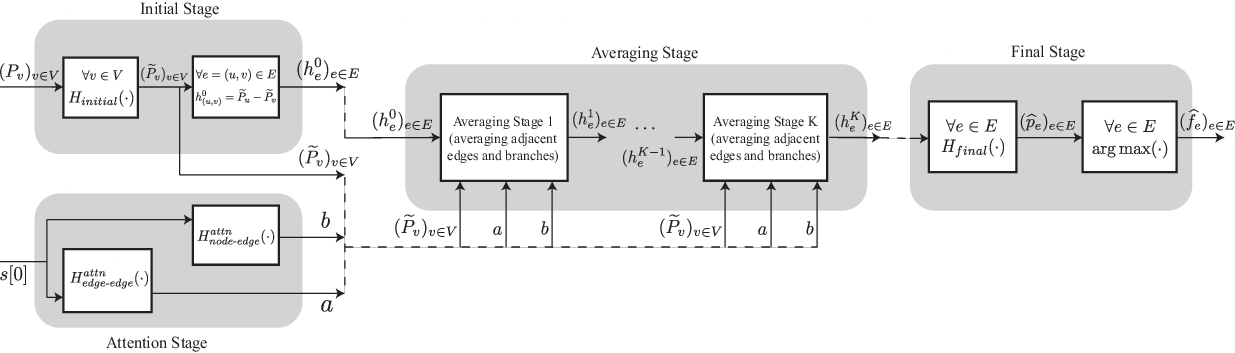}
	\caption{Block diagram of the GNN model. }
	\label{fig:blockDiagram}
\end{figure*}

The proposed model takes as input the topology of the grid $G = (V,E)$, the initial contingency $s[0] \in \{0,1\}^{|E|}$, and the power injection values $P_v \in \mathbb{R}$ at each node ${v\in V}$. In this model, we process the input data in multiple stages as explained in the following paragraphs. Fig. \ref{fig:blockDiagram} shows a block diagram that summarizes the model.

\subsubsection{Initial Stage} We start by removing the edges corresponding to failed branches in the initial contingency and get the new set of edges $E' = \{e\in E: s_{e}[0] = 1\}$. Then, we pass the node power injection values ${P}_v$ through a neural network to obtain the transformed node features $\widetilde{P}_v \in \mathbb{R}^L$ as follows.
\begin{equation}
	\forall v \in V, \quad \widetilde{P}_v = H_{initial}(P_v) 
\end{equation}
where, the mapping $H_{initial} : \mathbb{R} \rightarrow \mathbb{R}^L$ represents a dense neural network whose weights will be learned through back propagation during the training phase. Note that the same neural network is being used on all the nodes. Further, we use these values $\widetilde{P}_v$ to initiate the edge hidden features $h^0_e$ as 
\begin{equation}
		\forall e = (u,v) \in E', \quad h^0_e = h^0_{(u,v)} = \widetilde{P}_u - \widetilde{P}_v
\end{equation}
where, a directed edge $e\in E'$ is represented as $e=(u,v)$ with $u,v\in V$ being its source and destination nodes respectively. 

\subsubsection{Attention Stage} For an edge $e =(u,v) \in E'$, we define the set of adjacent edges as $\mathcal{N}_e= \mathcal{N}_{(u,v)} = \{(u,w) : w\in V,\; (u,w) \in E'  \}  \cup \{(w,v) : w\in V,\; (w,v)\in E' \}$. Now, we generate two types of attention coefficients: edge-to-edge $a_{ed}$ for every two neighboring edges $e\in E',d\in \mathcal{N}_e$; and node-to-edge $(b_{eu}, b_{ev})$ for all edges $e = (u,v) \in E'$ and their nodes $u,v$. We generate these coefficients by passing the given initial contingency $s[0]$ through two dense neural networks as
\begin{equation}
	a = H^{attn}_{edge\text{-}edge}(s[0]),\quad b = H^{attn}_{node\text{-}edge}(s[0])
\end{equation}
where, the mappings $H^{attn}_{edge\text{-}edge} : \RE^{|E|} \rightarrow \RE^{\sum_{e\in E'}\left(|\mathcal{N}_e| - 1\right)}$ and $H^{attn}_{node\text{-}edge} : \RE^{|E|} \rightarrow \RE^{2|E|}$ represent two dense neural networks whose weights will be learned through back propagation during training, $a$ is the collection of edge-to-edge coefficients $a = (a_{ed})_{\{e,d \in E' : d\in \mathcal{N}_e, d\neq e\}}$, and $b$ is the collection of node-to-edge coefficients $b = (b_{eu}, b_{ev})_{e=(u,v)\in E' }$.

These attention coefficients will be used in the next averaging stages, where we repetitively update edge hidden features by weighted-averages of neighboring edge and node hidden features. The attention coefficients will act as weights for this purpose. They represent how much weight, or attention, needs to be given on adjacent edges and nodes while updating an edge's hidden features.

\subsubsection{Averaging Stage} In this stage, we pass the outputs of the initial stage $h^0_e$ through  a sequence of $K$ averaging steps. In each step $k = 1,...,K$, we calculate the weighted average of neighboring edge and node features and pass them through a neural network to obtain the new edge features as follows.
\begin{multline}
	\forall k=1,...,K, \; \forall e=(u,v)\in E', \\
	h^k_e = \frac{h^{k-1}_e}{|\mathcal{N}_e|} + H^k_{edge\text{-}edge} \left( \sum_{d\in \mathcal{N}_e, d\neq e} \frac{a_{ed}}{\sqrt{|\mathcal{N}_e|} \sqrt{|\mathcal{N}_d|}} h_d^{k-1} \right) \\
	+ H^k_{node\text{-}edge} \left(\frac{b_{eu} \widetilde{P}_u + b_{ev}  \widetilde{P}_v}{2}\right)
\end{multline}
where, $K$ is the total number of averaging steps, $h_e^k$ is the edge hidden feature of edge $e$ at $k$-th averaging step, and $\mathcal{N}_e$ is the set of edges that are adjacent to and including $e$. The coefficients $a_{ed}$ and $b_{ue}$ are the edge-to-edge and node-to-edge attention coefficients respectively obtained in the attention stage. Finally, the functions $H^k_{edge\text{-}edge}, H^k_{node\text{-}edge} : \RE^L\rightarrow\RE^L$ for $k=1,...,K$ represent dense neural networks, whose weights will be learned through back propagation during the training phase. Note that these neural networks are the same for all edges.

\subsubsection{Final Stage} In the final stage, we predict the failure step probability values $\widehat{p}_e := [\widehat{p}_{e,0},...,\widehat{p}_{e,T-1}]$ for all branches $e\in E'$, where each entry $\widehat{p}_{e,t}$ is the predicted probability that branch $e$ fails at time $t$. We do this by passing the output of last averaging step $h_e^K$ through a dense neural network as
\begin{equation}
	\forall e \in E', \quad \widehat{p}_e = H_{final}(h_e^K) \in [0,1]^T
\end{equation}
where, the function $H_{final}:\RE^L \rightarrow \RE^T$ represents a dense neural network, whose weights will be learned during the training phase. This neural network has a softmax layer, so that its outputs represent valid probability values. Finally, the failure steps of edges is predicted by picking the index with the highest predicted probability value,
\begin{equation}
	\forall  e\in E', \quad \widehat{f}_e = \arg \max_{t=0,...,T-1} \widehat{p}_{e,t}.
\end{equation}

Having defined the model architecture, we explain methods to generate data for training and testing in the next section.

\section{Data Synthesis}\label{sec:datasynth}
We generate cascade data using the cascading failure simulator (CFS) oracle proposed in \cite{flowFreeRandomChem1}. Data generated from other oracles or historic data obtained from utility records  can also be used on our model without any changes to its architecture. 

\begin{algorithm}
	\caption{Simulating failure cascade using the CFS oracle.}\label{alg:cfs}
	\begin{algorithmic}
		\STATE \textbf{Input:} Grid topology $G = (V,E)$, initial contingency $s[0]$, capacity values $(c_e)_{e\in E}$, and power injection $(P_v)_{v\in V}$.
		\STATE \textbf{Output:} The branch failure steps in the cascade $(f_e)_{e\in E}$. 
		\STATE \textbf{initialize} $t \leftarrow 0;$ overloaded $\leftarrow$ true.
			\WHILE{overloaded}
			\STATE 1)\hspace{8pt}$E \leftarrow \{e\in E: s_{e}[t] = 1\}$.
			\STATE 2)\hspace{8pt}Detect the formed islands (disconnected sub-graphs). 
			\FOR{each island}
				\STATE a)\hspace{8pt}Specify a bus as the slack bus of the island.
				\STATE b)\hspace{7pt}Rebalance the power injection within the island.
				\STATE c)\hspace{8pt}Recompute the power flows $g_e$ within the island.
			\ENDFOR
			\STATE 3)\hspace{8pt}$s[t+1] \leftarrow s[t]; \quad \forall e\in E : g_e > c_e, \; s_e[t+1] \leftarrow 0$.
			\STATE 4)\hspace{8pt}\textbf{if} $s[t+1] = s[t]$ \textbf{then} overloaded $\leftarrow$ false \textbf{end if}
			\STATE 5)\hspace{8pt}$t \leftarrow t + 1$.
			\ENDWHILE
			\STATE \textbf{return} $f = \sum_t s[t]$.
	\end{algorithmic}
\end{algorithm}

As summarized in Algorithm \ref{alg:cfs}, the CFS oracle simulates the cascade process for a given initial contingency $s[0]$ and power injection values $(P_v)_{v\in V}$, and outputs the failure steps $f:=(f_e)_{e\in E}$. The CFS oracle treats branch failures deterministically, a branch $e$ is treated to be failed whenever the power flow $g_e$ through it crosses its given capacity value $c_e$. We implement the CFS oracle in MATLAB, where we use the MATPOWER toolbox \cite{matpower} to get the graph topology, default power injection values, and branch capacity values. We use its DC power flow solver to calculate the branch power flows.

Furthermore, using this CFS oracle, we generate a data pool $\mathcal{D}$ of size $M$ as summarized in Algorithm \ref{alg:data_gen}. Each sample in the data  pool $\mathcal{D}$ is a tuple $\left(s[0], (P_v)_{v\in V}, (f_e)_{e\in E}\right)$ containing a random initial contingency, randomly scaled power injection values, and the corresponding cascade failure steps respectively. Specifically, in each sample, we first generate random $|E|-2$ initial contingencies $s[0]$ by selecting two branches randomly, say $e_1,e_2$, and setting their states to failed $s_{e_1}[0] = s_{e_2}[0] = 0$. Then, we scale the default power injection values obtained from the MATPOWER toolbox uniformly across all nodes by a random scaling value $\alpha \sim \text{Unif}[1,2]$. We then use the CFS oracle described in Algorithm \ref{alg:cfs} to get the cascade failure steps $(f_e)_{e\in E}$ for each sample.
\begin{algorithm}
	\caption{Generating the failure cascade data pool.}\label{alg:data_gen}
	\begin{algorithmic}
		\STATE \textbf{Input:} Grid topology $G = (V,E)$, default node power injection values $(P_v^0)_{v\in V}$, and required data pool size $M$.
		\STATE \textbf{Output:} The failure cascade data pool $\mathcal{D}$.
		\STATE  \textbf{initialize} $\mathcal{D} \leftarrow \{\}$.
		\WHILE{$|\mathcal{D}| < M$}
		\STATE 1) $s[0] \leftarrow$ Random $|E|-2$ initial contingency.
		\STATE 2) $\alpha \leftarrow \text{Unif}[1,2]; \quad \forall v \in V, \; P_v \leftarrow \alpha P_v^0$.
		\STATE 3) $(f_e)_{e\in E} \leftarrow$ CFS oracle's output for $\left(s[0], (P_v)_{v\in V}, G\right)$.
		\STATE 4) $\mathcal{D} \leftarrow \mathcal{D} \text{ appended with } (s[0], (P_v)_{v\in V}, (f_e)_{e\in E})$.
		\ENDWHILE \STATE \textbf{return} $\mathcal{D}$.
	\end{algorithmic}
\end{algorithm}

In the next section, we discuss the details of model training with this data and present their performance results.


\section{Results}\label{sec:results}
We generate two data pools, one for IEEE89 system and another for IEEE118 system, each containing 200,000 cascade sequence samples simulated on random initial contingencies and random uniform scaling values. We get the required graph topology, default power injection values, and branch capacities (we set the unavailable capacities to twice the default power flows through the branches) from the MATPOWER toolbox. In both cases, we split the dataset $\mathcal{D}$ into train $\mathcal{D}_{train}$ (90\%) and test $\mathcal{D}_{test}$ (10\%) sets. We then train two instances of our proposed model, one for IEEE89 and another for IEEE118 systems\footnote{See implementations at https://github.com/sathwikchadaga/failure-cascade.}. We build the instances in Python using the PyTorch library \cite{pytorch}.
We train the models with an Adam optimizer and try to minimize the cross entropy loss between predicted and true failure steps averaged on all branches and batch samples. 

We use the performance of the influence model \cite{flowFreeInfluenceXinyu} as a benchmark since this model can evaluate metrics in almost the same granularity level as the GNN model. Note, however, that the influence model does not take node power injection values into consideration, making it specific to a single loading profile. Hence, it is impossible to generalize a single influence model over all load scaling values. Thus, for a given IEEE system, we build multiple influence model instances, each trained on a unique load scaling value, and compare our single generalized GNN model against them. The metrics we investigate can be categorized into two types: graph-level and branch-level metrics.


\subsubsection{Graph Level Metrics}
Here, we present several metrics, in increasing order of prediction granularity, that capture the graph-level prediction performance of the trained model.

\paragraph{Failure Size Error Rate}
The cascade failure size is defined as the number of branches in the failed state at the end of a cascade. Let $\mathcal{D}_{test}^\alpha \subset \mathcal{D}_{test}$ be the set of test samples whose injection load scaling is $\alpha$. For a sample $d \in \mathcal{D}^\alpha_{test}$ (with a random $|E|-2$ initial contingency), if $E_{failed}^d \subset E$ is the set of edges that have truly failed and $\widehat{E}_{failed}^d \subset E$ is the set of edges predicted to have failed, then the graph-level failure size error rate $l_{size}^\alpha$ at a load scaling value $\alpha$ is defined as $l_{size}^\alpha = \frac{1}{\left| \mathcal{D}^\alpha_{test} \right|}\sum_{d \in \mathcal{D}^\alpha_{test}} {||E_{failed}^d|-|\widehat{E}_{failed}^d||}/{|E^d_{failed}|}$.


\begin{figure}[htbp]\vspace{-6pt}
	\centering
	\resizebox {\columnwidth} {!} {
\begin{tikzpicture}

\definecolor{darkblue00167}{RGB}{0,0,167}
\definecolor{darkgray176}{RGB}{176,176,176}
\definecolor{firebrick1933945}{RGB}{193,39,45}
\definecolor{lightgray204}{RGB}{204,204,204}

\begin{groupplot}[group style={group size=2 by 1}]
\nextgroupplot[
legend cell align={left},
legend style={fill opacity=0.8, draw opacity=1, text opacity=1, draw=lightgray204},
tick align=outside,
tick pos=left,
x grid style={darkgray176},
xlabel={Load scaling},
xmin=0.923188644647598, xmax=2.07674996256828,
xtick style={color=black},
y grid style={darkgray176},
ylabel={Failure size error (\%)},
ymin=0, ymax=25,
ytick style={color=black}
]
\addplot [semithick, darkblue00167, dashed, mark=square*, mark size=2.25, mark options={solid}]
table {%
0.975623250007629 6.18811425798741
1.02770721912384 4.6403216243792
1.08042061328888 7.2755778930837
1.13228726387024 11.4201412381244
1.18471789360046 13.0685994990903
1.23787021636963 8.57926768440158
1.29045939445496 8.28063249606334
1.34462869167328 7.32901268032919
1.39903318881989 4.53260701839797
1.45040106773376 3.86929371176648
1.49875593185425 3.56483659978773
1.54848980903625 3.20978183355652
1.60126006603241 2.8007369543441
1.65564477443695 2.12021348468897
1.70725786685944 1.41308103715543
1.76134884357452 1.18880277408772
1.81366372108459 1.25835959701875
1.86528158187866 1.43140654153183
1.9159699678421 1.69276057336418
1.96697819232941 1.65802576511848
2.02188491821289 1.81419539195104
};
\addlegendentry{GNN (bin average)}
\addplot [semithick, firebrick1933945, mark=*, mark size=2.25, mark options={solid}, only marks]
table {%
1 18.1703003277861
1.1 21.9002683826847
1.2 55.1121049520729
1.3 11.9669462270099
1.4 12.2974206166242
1.5 6.43691210743357
1.6 6.24674009425519
1.7 4.37986229434396
1.8 3.89351408355902
1.9 4.36296550746219
2 4.61722594915359
};
\addlegendentry{Influence models}

\nextgroupplot[
legend cell align={left},
legend style={
  fill opacity=0.8,
  draw opacity=1,
  text opacity=1,
  at={(0.91,0.5)},
  anchor=east,
  draw=lightgray204
},
tick align=outside,
tick pos=left,
x grid style={darkgray176},
xlabel={Load scaling},
xmin=0.923188644647598, xmax=2.07674996256828,
xtick style={color=black},
y grid style={darkgray176},
ymin=0.649255039166676, ymax=3.60837593765405,
ytick style={color=black}
]
\addplot [semithick, darkblue00167, dashed, mark=square*, mark size=2.25, mark options={solid}]
table {%
0.977427005767822 1.86742451115221
1.03001689910889 1.03762091351908
1.08261823654175 1.07160777247465
1.13787531852722 1.17841572890512
1.19125521183014 1.36170378360958
1.24404168128967 1.24706387908018
1.29405379295349 1.25230760444872
1.34606111049652 1.10250290896462
1.39747321605682 1.24092515857249
1.44819092750549 1.21733433202285
1.499302983284 1.27184170545613
1.55119824409485 1.37369778382928
1.6031905412674 0.879706867558302
1.65577554702759 0.783760534552466
1.70964622497559 1.1065848750047
1.75970923900604 1.30357808306688
1.81061780452728 1.00442238176455
1.86352777481079 0.885064538474343
1.91665089130402 1.02143328123894
1.97011077404022 1.01439423151627
2.02431535720825 0.964136839462597
};
\addlegendentry{GNN (bin average)}
\addplot [semithick, firebrick1933945, mark=*, mark size=2.25, mark options={solid}, only marks]
table {%
1 3.10062962222461
1.1 2.43076997439411
1.2 3.20284673669159
1.3 2.45816010202442
1.4 2.78759674174001
1.5 3.08825534649397
1.6 3.07051213977265
1.7 3.47387044226826
1.8 3.22723149639525
1.9 2.93528972848821
2 2.81851253142877
};
\addlegendentry{Influence models}
\end{groupplot}

\end{tikzpicture} }
	\caption{Failure size error rates $l_{size}^\alpha$ of various models for IEEE89 (left) and IEEE118 (right) against scaling values $\alpha$.}
	\label{fig:sampleMetricFailureSize}
\end{figure}
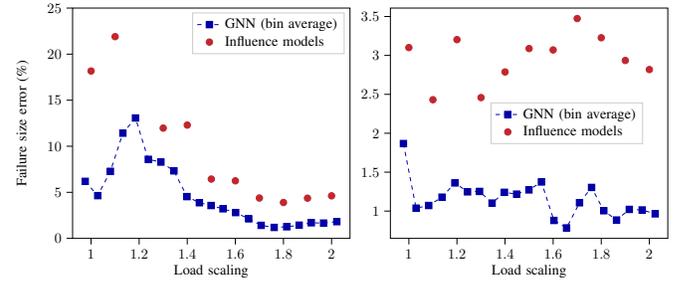
Fig. \ref{fig:sampleMetricFailureSize} shows the failure size error rates of two instances of the GNN model trained and tested on IEEE89 (left) and IEEE118 (right) cases respectively, along with the error rates of multiple influence model instances specific to the corresponding scaling value (recall the influence model's inability to generalize over variable load profiles). 
As can be seen, even though we are comparing the same generic GNN model to multiple load-specific influence models, the GNN model has lower error rates at all loading values. 

\paragraph{Final State Error Rate} 
If the true final state of a cascade sample $d \in \mathcal{D}^\alpha_{test}$  is $s^d[T] = (s^d_e[T])_{e\in E}$, and its predicted final state is $\hat{s}^d[T] = (\hat{s}^d_e[T])_{e\in E}$, then the graph-level final state error rate $l_{state}^\alpha$ at a load scaling value $\alpha$ is  defined as $l_{state}^\alpha =   \frac{1}{\left| \mathcal{D}^\alpha_{test} \right|}\sum_{d \in \mathcal{D}^\alpha_{test}} \frac{1}{|E|}\sum_{e\in E} \left|s^d_e[T] - \hat{s}^d_e[T]\right|$.


\begin{figure}[htbp]
	\centering
	\resizebox {\columnwidth} {!} {
\begin{tikzpicture}

\definecolor{darkblue00167}{RGB}{0,0,167}
\definecolor{darkgray176}{RGB}{176,176,176}
\definecolor{firebrick1933945}{RGB}{193,39,45}
\definecolor{lightgray204}{RGB}{204,204,204}

\begin{groupplot}[group style={group size=2 by 1}]
\nextgroupplot[
legend cell align={left},
legend style={fill opacity=0.8, draw opacity=1, text opacity=1, draw=lightgray204},
tick align=outside,
tick pos=left,
x grid style={darkgray176},
xlabel={Load scaling},
xmin=0.923188644647598, xmax=2.07674996256828,
xtick style={color=black},
y grid style={darkgray176},
ylabel={Final state error (\%)},
ymin=0.43598230140389, ymax=7.14424124963264,
ytick style={color=black}
]
\addplot [semithick, darkblue00167, dashed, mark=square*, mark size=2.25, mark options={solid}]
table {%
0.975623250007629 0.970376424628814
1.02770721912384 0.740903162687015
1.08042061328888 1.42597089132264
1.13228726387024 2.53069677228938
1.18471789360046 4.02450067812546
1.23787021636963 4.85181942845092
1.29045939445496 4.56191316399662
1.34462869167328 4.6590724331679
1.39903318881989 2.742768223886
1.45040106773376 2.38690139594324
1.49875593185425 2.32621759919786
1.54848980903625 1.85353270677083
1.60126006603241 1.66769459660425
1.65564477443695 1.32730277997105
1.70725786685944 0.828873556092015
1.76134884357452 0.764869066377898
1.81366372108459 0.82738143854515
1.86528158187866 0.922432088927061
1.9159699678421 1.08526978705872
1.96697819232941 0.991270310769812
2.02188491821289 1.18847486442413
};
\addlegendentry{GNN (bin average)}
\addplot [semithick, firebrick1933945, mark=*, mark size=2.25, mark options={solid}, only marks]
table {%
1 2.01407766990291
1.1 4.03252427184466
1.2 6.83932038834952
1.3 5.59660194174757
1.4 4.93543689320388
1.5 4.26067961165049
1.6 4.05679611650485
1.7 3.07281553398058
1.8 2.95922330097087
1.9 3.18495145631068
2 3.1252427184466
};
\addlegendentry{Influence models}

\nextgroupplot[
legend cell align={left},
legend style={
  fill opacity=0.8,
  draw opacity=1,
  text opacity=1,
  at={(0.91,0.5)},
  anchor=east,
  draw=lightgray204
},
tick align=outside,
tick pos=left,
x grid style={darkgray176},
xlabel={Load scaling},
xmin=0.923188644647598, xmax=2.07674996256828,
xtick style={color=black},
y grid style={darkgray176},
ymin=0.638554032830435, ymax=3.76092014581627,
ytick style={color=black}
]
\addplot [semithick, darkblue00167, dashed, mark=square*, mark size=2.25, mark options={solid}]
table {%
0.977427005767822 1.41381990219482
1.03001689910889 0.861461875318479
1.08261823654175 0.876053181828045
1.13787531852722 0.953593698118915
1.19125521183014 1.23981129970463
1.24404168128967 1.2675460847486
1.29405379295349 1.13076972173618
1.34606111049652 1.06802493884784
1.39747321605682 1.19487815631218
1.44819092750549 1.19243224152998
1.499302983284 1.19659534511996
1.55119824409485 1.37610906049484
1.6031905412674 0.909252190687617
1.65577554702759 0.780479765238882
1.70964622497559 1.11188063098759
1.75970923900604 1.32094733279292
1.81061780452728 1.12704914283068
1.86352777481079 0.981174560905505
1.91665089130402 1.07839542303662
1.97011077404022 0.962396099286921
2.02431535720825 0.901238636525928
};
\addlegendentry{GNN (bin average)}
\addplot [semithick, firebrick1933945, mark=*, mark size=2.25, mark options={solid}, only marks]
table {%
1 3.08324022346369
1.1 2.96424581005587
1.2 3.61899441340782
1.3 3.18715083798883
1.4 3.59944134078212
1.5 3.56480446927374
1.6 3.23072625698324
1.7 3.25083798882682
1.8 3.41731843575419
1.9 3.19832402234637
2 3.20223463687151
};
\addlegendentry{Influence models}
\end{groupplot}

\end{tikzpicture} }
	\caption{Final state error rates $l_{state}^\alpha$ of various models for IEEE89 (left) and IEEE118 (right) against scaling values $\alpha$.}
	\label{fig:sampleMetricFinalState}\vspace{-4pt}
\end{figure}
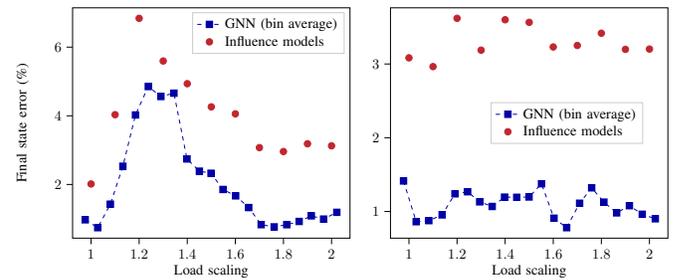

Fig. \ref{fig:sampleMetricFinalState} shows the final state error rates of two instances of the GNN model for IEEE89 (left) and IEEE118 (right) systems respectively, along with the error rates of different load-specific influence models. As can be seen, the GNN model is better by around 2\% (nearly a factor of 2) than all the load-specific influence models. 

\paragraph{Failure Step Error Rate} 
In a cascade sample $d \in \mathcal{D}^\alpha_{test}$, if the true failure steps are $f^d = (f^d_e)_{e\in E}$ and the predicted failure steps are $\widehat{f}^d = (\widehat{f}^d_e)_{e\in E}$, then the graph-level failure step error $l_{failure{-}step}^\alpha$ at a load scaling $\alpha$ is defined as $l_{failure{-}step}^\alpha = \frac{1}{\left| \mathcal{D}^\alpha_{test} \right|}\sum_{d \in \mathcal{D}^\alpha_{test}} \frac{1}{|E|} \sum_{e\in E} |f^d_e - \widehat{f}^d_e|$.

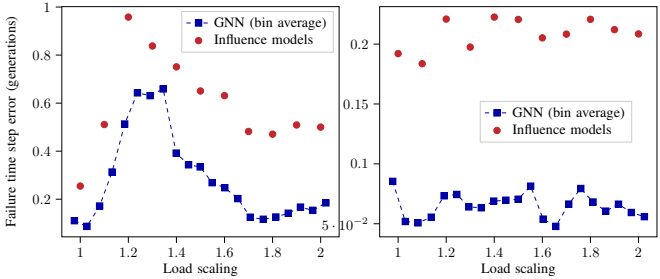
\begin{figure}[htbp]\vspace{-6pt}
	\centering
	\resizebox {\columnwidth} {!} {
\begin{tikzpicture}

\definecolor{darkblue00167}{RGB}{0,0,167}
\definecolor{darkgray176}{RGB}{176,176,176}
\definecolor{firebrick1933945}{RGB}{193,39,45}
\definecolor{lightgray204}{RGB}{204,204,204}

\begin{groupplot}[group style={group size=2 by 1}]
\nextgroupplot[
legend cell align={left},
legend style={fill opacity=0.8, draw opacity=1, text opacity=1, draw=lightgray204},
tick align=outside,
tick pos=left,
x grid style={darkgray176},
xlabel={Load scaling},
xmin=0.923188644647598, xmax=2.07674996256828,
xtick style={color=black},
y grid style={darkgray176},
ylabel={Failure time step error (generations)},
ymin=0.043390306512205, ymax=1.00171049395712,
ytick style={color=black}
]
\addplot [semithick, darkblue00167, dashed, mark=square*, mark size=2.25, mark options={solid}]
table {%
0.975623250007629 0.110665686015845
1.02770721912384 0.0869503150324282
1.08042061328888 0.171586524560612
1.13228726387024 0.313162889368754
1.18471789360046 0.513100832211212
1.23787021636963 0.642928535885671
1.29045939445496 0.631217178250243
1.34462869167328 0.659245743594297
1.39903318881989 0.391982834284732
1.45040106773376 0.343370118500281
1.49875593185425 0.334533665222345
1.54848980903625 0.26892796202324
1.60126006603241 0.247876216072043
1.65564477443695 0.20269540089433
1.70725786685944 0.125049738798836
1.76134884357452 0.11651505372084
1.81366372108459 0.125952472437944
1.86528158187866 0.141735336601445
1.9159699678421 0.167080317998446
1.96697819232941 0.153728483611632
2.02188491821289 0.184968369229998
};
\addlegendentry{GNN (bin average)}
\addplot [semithick, firebrick1933945, mark=*, mark size=2.25, mark options={solid}, only marks]
table {%
1 0.254708737864078
1.1 0.510922330097087
1.2 0.958150485436893
1.3 0.837859223300971
1.4 0.750956310679612
1.5 0.650529126213592
1.6 0.631155339805825
1.7 0.482043689320388
1.8 0.470776699029126
1.9 0.509097087378641
2 0.500165048543689
};
\addlegendentry{Influence models}

\nextgroupplot[
legend cell align={left},
legend style={
  fill opacity=0.8,
  draw opacity=1,
  text opacity=1,
  at={(0.91,0.5)},
  anchor=east,
  draw=lightgray204
},
tick align=outside,
tick pos=left,
x grid style={darkgray176},
xlabel={Load scaling},
xmin=0.923188644647598, xmax=2.07674996256828,
xtick style={color=black},
y grid style={darkgray176},
ymin=0.0390724100711024, ymax=0.231272689171927,
ytick style={color=black}
]
\addplot [semithick, darkblue00167, dashed, mark=square*, mark size=2.25, mark options={solid}]
table {%
0.977427005767822 0.0853815563392852
1.03001689910889 0.0518931013364697
1.08261823654175 0.0507601408107057
1.13787531852722 0.0554668772183046
1.19125521183014 0.0733211576920976
1.24404168128967 0.0744507270674378
1.29405379295349 0.0640769505541253
1.34606111049652 0.0633655672035955
1.39747321605682 0.0688936690839015
1.44819092750549 0.0696328783543788
1.499302983284 0.0704734838979422
1.55119824409485 0.0813752845513728
1.6031905412674 0.053853945194965
1.65577554702759 0.0478087863938672
1.70964622497559 0.0664180534160372
1.75970923900604 0.0794211510131566
1.81061780452728 0.0681495534761362
1.86352777481079 0.0605253254453761
1.91665089130402 0.0663236078809084
1.97011077404022 0.0593810126962172
2.02431535720825 0.0559174352808336
};
\addlegendentry{GNN (bin average)}
\addplot [semithick, firebrick1933945, mark=*, mark size=2.25, mark options={solid}, only marks]
table {%
1 0.192061452513966
1.1 0.183620111731844
1.2 0.220882681564246
1.3 0.197480446927374
1.4 0.222536312849162
1.5 0.220575418994413
1.6 0.205201117318436
1.7 0.208307262569832
1.8 0.220675977653631
1.9 0.212061452513966
2 0.208469273743017
};
\addlegendentry{Influence models}
\end{groupplot}

\end{tikzpicture} }
	\caption{Failure step error rates $l_{failure\text{-}step}^\alpha$ of various models for IEEE89 (left) and IEEE118 (right) against load scaling $\alpha$.}
	\label{fig:sampleMetricFailureStep}\vspace{-6pt}
\end{figure}

Fig. \ref{fig:sampleMetricFailureStep} shows the failure step error rates for two instances of the GNN models for IEEE89 (left) and IEEE118 (right) systems respectively, along with the error rates of different load-specific influence models. 
For both buses, the generic GNN model outperforms all the load-specific influence models.


\subsubsection{Branch Level Metrics}
In this subsection, we present several branch metrics that depict the model's accuracy in prediction of branch features. But first, we define the \emph{branch failure frequency} $l_{freq,e}$ for any branch $e\in E$. Let $\mathcal{D}^e_{failed} \subset \mathcal{D}_{train}$ be the set of train samples where the branch has eventually failed in the cascade but not as part of the initial contingency, then we define $l_{freq,e} = {|\mathcal{D}^e_{failed}|}/{|\mathcal{D}_{train}|}$.
This value captures the prediction difficulty of  branches. For example, predicting the features of a branch that rarely fails is easier than a branch that fails half the time.

\paragraph{Branch Final State Error Rate}

Let $\mathcal{D}_{wrong}^e \subset \mathcal{D}_{test}$ be the set of test samples in which the model wrongly predicted the final state of edge $e \in E$. Further, let $\mathcal{D}_{initial}^e \subset \mathcal{D}_{wrong}^e$ be the samples in which branch $e$ failed as part of the initial contingency. We do not count such samples as predicting their states is trivial, hence we define the branch final state error as the ratio $l_{state,e} = |\mathcal{D}_{wrong}^e|/(|\mathcal{D}_{test}|-|\mathcal{D}_{initial}^e|)$.

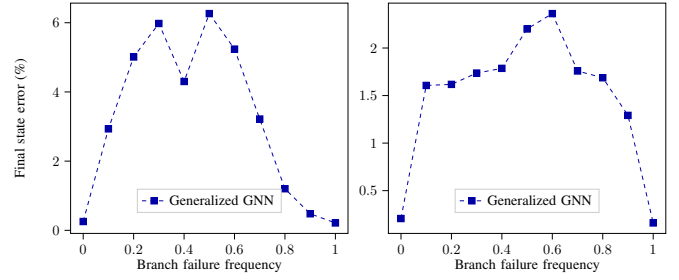
\begin{figure}[htbp]\vspace{-8pt}
	\centering
	\resizebox {\columnwidth} {!} {
\begin{tikzpicture}

\definecolor{darkblue00167}{RGB}{0,0,167}
\definecolor{darkgray176}{RGB}{176,176,176}
\definecolor{lightgray204}{RGB}{204,204,204}

\begin{groupplot}[group style={group size=2 by 1}]
\nextgroupplot[
legend cell align={left},
legend style={
  fill opacity=0.8,
  draw opacity=1,
  text opacity=1,
  at={(0.5,0.09)},
  anchor=south,
  draw=lightgray204
},
tick align=outside,
tick pos=left,
x grid style={darkgray176},
xlabel={Branch failure frequency},
xmin=-0.05, xmax=1.05,
xtick style={color=black},
y grid style={darkgray176},
ylabel={Final state error (\%)},
ymin=-0.0892735376164079, ymax=6.5666007350563,
ytick style={color=black}
]
\addplot [semithick, darkblue00167, dashed, mark=square*, mark size=2.25, mark options={solid}]
table {%
0 0.254206978246916
0.1 2.93270191727179
0.2 5.00988504090491
0.3 5.97788171721597
0.4 4.30198601778106
0.5 6.26406099538936
0.6 5.23833159713075
0.7 3.21679601219097
0.8 1.20189840212834
0.9 0.478344271341908
1 0.213266202050534
};
\addlegendentry{Generalized GNN}

\nextgroupplot[
legend cell align={left},
legend style={
  fill opacity=0.8,
  draw opacity=1,
  text opacity=1,
  at={(0.5,0.09)},
  anchor=south,
  draw=lightgray204
},
tick align=outside,
tick pos=left,
x grid style={darkgray176},
xlabel={Branch failure frequency},
xmin=-0.05, xmax=1.05,
xtick style={color=black},
y grid style={darkgray176},
ymin=0.050170140560149, ymax=2.47211936377038,
ytick style={color=black}
]
\addplot [semithick, darkblue00167, dashed, mark=square*, mark size=2.25, mark options={solid}]
table {%
0 0.206673089607727
0.1 1.60639570731799
0.2 1.61754602242174
0.3 1.73643043501922
0.4 1.78610574270838
0.5 2.20031527656576
0.6 2.36203076271537
0.7 1.75846138399398
0.8 1.68711674925956
0.9 1.29150696135372
1 0.160258741615159
};
\addlegendentry{Generalized GNN}
\end{groupplot}

\end{tikzpicture} }
	\caption{Branch final state prediction error rate $l_{state,e}$ for IEEE89 (left) and IEEE118 (right) averaged over all load scaling values $[1,2]$ against branch failure frequencies $l_{freq,e}$.}
	\label{fig:branchMetricFinalStaterandom}\vspace{-8pt}
\end{figure}

Fig. \ref{fig:branchMetricFinalStaterandom} shows the final state error rate for two instances of the GNN model trained and tested on IEEE89 (left) and IEEE118 (right) datasets respectively. The error rates are averaged over all test samples containing random initial contingencies and random load scaling values. As can be seen, the branch final state prediction error rate by the GNN model is below 6\% and 2.5\% at all branches for IEEE89 and IEEE118 systems respectively. 

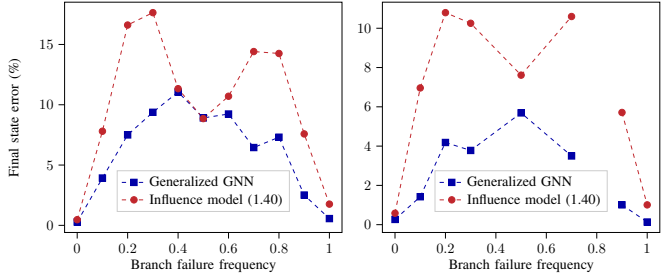
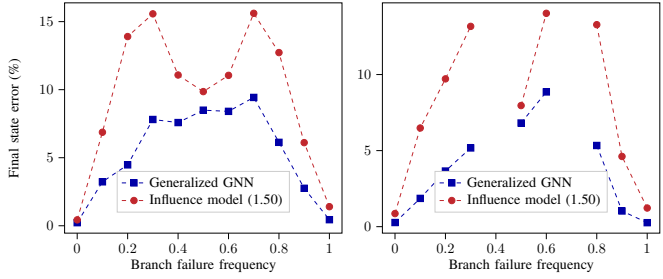
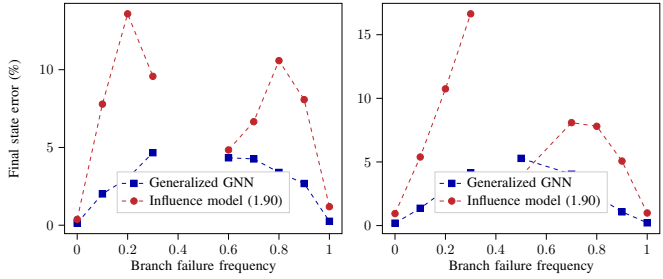
\begin{figure}[htbp]
	\centering
	\begin{subfigure}[b]{\columnwidth}
		\centering
		\resizebox {\columnwidth} {!} {
\begin{tikzpicture}

\definecolor{darkblue00167}{RGB}{0,0,167}
\definecolor{darkgray176}{RGB}{176,176,176}
\definecolor{firebrick1933945}{RGB}{193,39,45}
\definecolor{lightgray204}{RGB}{204,204,204}

\begin{groupplot}[group style={group size=2 by 1}]
\nextgroupplot[
legend cell align={left},
legend style={
  fill opacity=0.8,
  draw opacity=1,
  text opacity=1,
  at={(0.5,0.09)},
  anchor=south,
  draw=lightgray204
},
tick align=outside,
tick pos=left,
x grid style={darkgray176},
xlabel={Branch failure frequency},
xmin=-0.05, xmax=1.05,
xtick style={color=black},
y grid style={darkgray176},
ylabel={Final state error (\%)},
ymin=-0.590377823368007, ymax=18.5058913063033,
ytick style={color=black}
]
\addplot [semithick, darkblue00167, dashed, mark=square*, mark size=2.25, mark options={solid}]
table {%
0 0.27763440979887
0.1 3.92444418642435
0.2 7.49875812991456
0.3 9.37544812165114
0.4 11.053789558987
0.5 8.90896092668114
0.6 9.23204460191275
0.7 6.45249103664472
0.8 7.30953572473142
0.9 2.51439669250665
1 0.561040766206518
};
\addlegendentry{Generalized GNN}
\addplot [semithick, firebrick1933945, dashed, mark=*, mark size=2.25, mark options={solid}]
table {%
0 0.469086903477368
0.1 7.80405899259617
0.2 16.6104442468489
0.3 17.6378790731364
0.4 11.3324645231086
0.5 8.85240781976517
0.6 10.7054132494888
0.7 14.4113583475994
0.8 14.2535960084974
0.9 7.58253070623091
1 1.76075650660523
};
\addlegendentry{Influence model (1.40)}

\nextgroupplot[
legend cell align={left},
legend style={
  fill opacity=0.8,
  draw opacity=1,
  text opacity=1,
  at={(0.5,0.09)},
  anchor=south,
  draw=lightgray204
},
tick align=outside,
tick pos=left,
unbounded coords=jump,
x grid style={darkgray176},
xlabel={Branch failure frequency},
xmin=-0.05, xmax=1.05,
xtick style={color=black},
y grid style={darkgray176},
ymin=-0.397197969733797, ymax=11.3275026940815,
ytick style={color=black}
]
\addplot [semithick, darkblue00167, dashed, mark=square*, mark size=2.25, mark options={solid}]
table {%
0 0.266788175033686
0.1 1.4217120836174
0.2 4.18436976499843
0.3 3.77918347276182
0.5 5.69028126596867
0.7 3.49828017510944
0.8 nan
0.9 1.00848653923768
1 0.135742969530536
};
\addlegendentry{Generalized GNN}
\addplot [semithick, firebrick1933945, dashed, mark=*, mark size=2.25, mark options={solid}]
table {%
0 0.58336825442445
0.1 6.96055885363078
0.2 10.7945617548172
0.3 10.2530673422474
0.5 7.61263685486589
0.7 10.5979065863533
0.8 nan
0.9 5.71219828797086
1 1.00744618297557
};
\addlegendentry{Influence model (1.40)}
\end{groupplot}

\end{tikzpicture} }
		\caption{Load scaling = $1.40$ for IEEE89 (left) and IEEE118 (right).}
	\end{subfigure}\vspace{6pt}
	\begin{subfigure}[b]{\columnwidth}
		\centering
		\resizebox {\columnwidth} {!} {
\begin{tikzpicture}

\definecolor{darkblue00167}{RGB}{0,0,167}
\definecolor{darkgray176}{RGB}{176,176,176}
\definecolor{firebrick1933945}{RGB}{193,39,45}
\definecolor{lightgray204}{RGB}{204,204,204}

\begin{groupplot}[group style={group size=2 by 1}]
\nextgroupplot[
legend cell align={left},
legend style={
  fill opacity=0.8,
  draw opacity=1,
  text opacity=1,
  at={(0.5,0.09)},
  anchor=south,
  draw=lightgray204
},
tick align=outside,
tick pos=left,
x grid style={darkgray176},
xlabel={Branch failure frequency},
xmin=-0.05, xmax=1.05,
xtick style={color=black},
y grid style={darkgray176},
ylabel={Final state error (\%)},
ymin=-0.542578026352178, ymax=16.3797557593484,
ytick style={color=black}
]
\addplot [semithick, darkblue00167, dashed, mark=square*, mark size=2.25, mark options={solid}]
table {%
0 0.226618963906941
0.1 3.2291172454241
0.2 4.47492446434168
0.3 7.80900157446787
0.4 7.58281728383312
0.5 8.49236879720444
0.6 8.40627259918334
0.7 9.42966335092728
0.8 6.12932374024557
0.9 2.75334081362734
1 0.44655891087436
};
\addlegendentry{Generalized GNN}
\addplot [semithick, firebrick1933945, dashed, mark=*, mark size=2.25, mark options={solid}]
table {%
0 0.432865922321277
0.1 6.86956526431398
0.2 13.9026863344349
0.3 15.5716446795284
0.4 11.0774434182387
0.5 9.85806984522651
0.6 11.0508877142966
0.7 15.6105587690893
0.8 12.7254042308709
0.9 6.1082415859398
1 1.40128795420461
};
\addlegendentry{Influence model (1.50)}

\nextgroupplot[
legend cell align={left},
legend style={
  fill opacity=0.8,
  draw opacity=1,
  text opacity=1,
  at={(0.5,0.09)},
  anchor=south,
  draw=lightgray204
},
tick align=outside,
tick pos=left,
unbounded coords=jump,
x grid style={darkgray176},
xlabel={Branch failure frequency},
xmin=-0.05, xmax=1.05,
xtick style={color=black},
y grid style={darkgray176},
ymin=-0.425563282019628, ymax=14.7131448902592,
ytick style={color=black}
]
\addplot [semithick, darkblue00167, dashed, mark=square*, mark size=2.25, mark options={solid}]
table {%
0 0.272252125733916
0.1 1.85855542715988
0.2 3.65675959838564
0.3 5.17529215401602
0.4 nan
0.5 6.81450716983712
0.6 8.85735527867508
0.7 nan
0.8 5.33652843621145
0.9 1.03060717151522
1 0.262559816720319
};
\addlegendentry{Generalized GNN}
\addplot [semithick, firebrick1933945, dashed, mark=*, mark size=2.25, mark options={solid}]
table {%
0 0.861145796967103
0.1 6.48542837285138
0.2 9.71513058736279
0.3 13.1667744147023
0.4 nan
0.5 7.96168110888923
0.6 14.0250217915193
0.7 nan
0.8 13.2779606953742
0.9 4.61583356641647
1 1.22551411383271
};
\addlegendentry{Influence model (1.50)}
\end{groupplot}

\end{tikzpicture} }
		\caption{Load scaling = $1.50$ for IEEE89 (left) and IEEE118 (right).}
	\end{subfigure}\vspace{6pt}
	\begin{subfigure}[b]{\columnwidth}
		\centering
		\resizebox {\columnwidth} {!} {
\begin{tikzpicture}

\definecolor{darkblue00167}{RGB}{0,0,167}
\definecolor{darkgray176}{RGB}{176,176,176}
\definecolor{firebrick1933945}{RGB}{193,39,45}
\definecolor{lightgray204}{RGB}{204,204,204}

\begin{groupplot}[group style={group size=2 by 1}]
\nextgroupplot[
legend cell align={left},
legend style={
  fill opacity=0.8,
  draw opacity=1,
  text opacity=1,
  at={(0.5,0.09)},
  anchor=south,
  draw=lightgray204
},
tick align=outside,
tick pos=left,
unbounded coords=jump,
x grid style={darkgray176},
xlabel={Branch failure frequency},
xmin=-0.05, xmax=1.05,
xtick style={color=black},
y grid style={darkgray176},
ylabel={Final state error (\%)},
ymin=-0.541980270322836, ymax=14.2647928329069,
ytick style={color=black}
]
\addplot [semithick, darkblue00167, dashed, mark=square*, mark size=2.25, mark options={solid}]
table {%
0 0.131054870733062
0.1 2.01452605683773
0.2 3.01362399644853
0.3 4.66298100522295
0.4 nan
0.5 nan
0.6 4.33595776496079
0.7 4.26235811779832
0.8 3.39205469403869
0.9 2.68059588531038
1 0.251168028324637
};
\addlegendentry{Generalized GNN}
\addplot [semithick, firebrick1933945, dashed, mark=*, mark size=2.25, mark options={solid}]
table {%
0 0.377163393202784
0.1 7.78580100935829
0.2 13.591757691851
0.3 9.57341328509498
0.4 nan
0.5 nan
0.6 4.83978760413806
0.7 6.65523370947954
0.8 10.580685120865
0.9 8.07393097917366
1 1.1913512985985
};
\addlegendentry{Influence model (1.90)}

\nextgroupplot[
legend cell align={left},
legend style={
  fill opacity=0.8,
  draw opacity=1,
  text opacity=1,
  at={(0.5,0.09)},
  anchor=south,
  draw=lightgray204
},
tick align=outside,
tick pos=left,
unbounded coords=jump,
x grid style={darkgray176},
xlabel={Branch failure frequency},
xmin=-0.05, xmax=1.05,
xtick style={color=black},
y grid style={darkgray176},
ymin=-0.6352155042404, ymax=17.4653293707694,
ytick style={color=black}
]
\addplot [semithick, darkblue00167, dashed, mark=square*, mark size=2.25, mark options={solid}]
table {%
0 0.187536535532773
0.1 1.35782269427273
0.2 2.79847894706539
0.3 4.14562137110723
0.4 nan
0.5 5.28171092793286
0.7 4.03453749927253
0.8 2.30107079099006
0.9 1.07962336260022
1 0.217287920410149
};
\addlegendentry{Generalized GNN}
\addplot [semithick, firebrick1933945, dashed, mark=*, mark size=2.25, mark options={solid}]
table {%
0 0.941505546079238
0.1 5.38781561431034
0.2 10.7446300530282
0.3 16.6425773309962
0.4 nan
0.5 3.90633541533781
0.7 8.09013057892489
0.8 7.80154004005481
0.9 5.07086310175735
1 0.989771909218337
};
\addlegendentry{Influence model (1.90)}
\end{groupplot}

\end{tikzpicture} }
		\caption{Load scaling = $1.90$ for IEEE89 (left) and IEEE118 (right)s.}
	\end{subfigure}
	\caption{Branch final state prediction error rate $l_{state,e}$ for IEEE89 (left) and IEEE118 (right) for various load scaling plotted against branch failure frequencies $l_{freq,e}$.}
	\label{fig:branchFinalStates}\vspace{-8pt}
\end{figure}

Further, the error plot in Fig. \ref{fig:branchMetricFinalStaterandom} is generated by averaging over random scaling values in $[1,2]$, which demonstrates that the GNN model can be generalized over variable load profiles. However, note that the plot does not contain the performance of a reference influence model, again because of the influence model's inability to generalize over variable load profiles. 
Hence, in order to benchmark the performance, Fig. \ref{fig:branchFinalStates} plots the final state error rates of the same GNN model, tested on three example load scaling values of $1.40$, $1.50$, and $1.90$, against different instances of the influence model built specifically for those load scaling values of $1.40$, $1.50$, and $1.90$.

As seen from these plots, the generic GNN model outperforms the load-specific instances of the influence model at almost all branches, beating the influence model by almost 10\% in some cases. However, the influence model performs better than the GNN model at some branches with failure frequencies close to 0.5.

\paragraph{Branch Failure Step Error Rate} 
Let $\mathcal{D}^e_{failed} \subset \mathcal{D}_{test}$ be the set of test samples where the branch $e \in E$ has eventually failed in the cascade but not as part of the initial contingency. Say, the true failure step of the branch $e$ in sample $d\in \mathcal{D}^e_{failed}$ is $f_e^d$ and the predicted state is  $\widehat{f}_e^d$, then we define the branch failure step error rate as $l_{failure{-}step,e} = \frac{1}{|\mathcal{D}^e_{failed}|}  \sum_{d\in \mathcal{D}^e_{failed}}|\widehat{f}_e^d - {f}_e^d|$.

\begin{figure}[htbp]
	\centering
	\resizebox {\columnwidth} {!} {
\begin{tikzpicture}

\definecolor{darkblue00167}{RGB}{0,0,167}
\definecolor{darkgray176}{RGB}{176,176,176}
\definecolor{lightgray204}{RGB}{204,204,204}

\begin{groupplot}[group style={group size=2 by 1}]
\nextgroupplot[
legend cell align={left},
legend style={fill opacity=0.8, draw opacity=1, text opacity=1, draw=lightgray204},
tick align=outside,
tick pos=left,
x grid style={darkgray176},
xlabel={Branch failure frequency},
xmin=-0.05, xmax=1.05,
xtick style={color=black},
y grid style={darkgray176},
ylabel={Failure time step error (generations)},
ymin=-0.00169346338718585, ymax=0.0705618078852736,
ytick style={color=black}
]
\addplot [semithick, darkblue00167, dashed, mark=square*, mark size=2.25, mark options={solid}]
table {%
0 0.01327343700848
0.1 0.0251348879381937
0.2 0.0608021823818886
0.3 0.0672774773728891
0.4 0.0400765989525269
0.5 0.0447005082737262
0.6 0.0389604735162111
0.7 0.045382345578686
0.8 0.0364977471534128
0.9 0.00432888676185387
1 0.00159086712519867
};
\addlegendentry{Generalized GNN}

\nextgroupplot[
legend cell align={left},
legend style={fill opacity=0.8, draw opacity=1, text opacity=1, draw=lightgray204},
tick align=outside,
tick pos=left,
x grid style={darkgray176},
xlabel={Branch failure frequency},
xmin=-0.05, xmax=1.05,
xtick style={color=black},
y grid style={darkgray176},
ymin=9.59907432375637e-05, ymax=0.0383901539618475,
ytick style={color=black}
]
\addplot [semithick, darkblue00167, dashed, mark=square*, mark size=2.25, mark options={solid}]
table {%
0 0.0366495101791834
0.1 0.00964720116928654
0.2 0.00315712863843757
0.3 0.00386173367432842
0.4 0.0153049172050258
0.5 0.010690809913724
0.6 0.0076300355447403
0.7 0.00389172502709736
0.8 0.00298741220751503
0.9 0.00183663452590165
1 0.00213257853523307
};
\addlegendentry{Generalized GNN}
\end{groupplot}

\end{tikzpicture} }
	\caption{Branch failure steps prediction error $l_{failure{-}step,e}$ for IEEE89 (left) and IEEE118 (right) averaged over all scaling $[1,2]$ against failure frequencies $l_{freq,e}$.}
	\label{fig:branchMetricFailureSteprandom}
\end{figure}
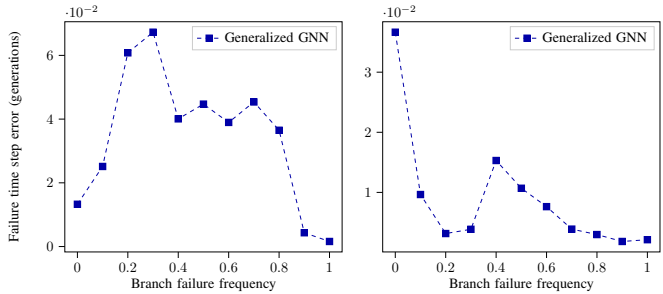

Fig. \ref{fig:branchMetricFailureSteprandom} shows the branch failure step error rate $l_{failure{-}step,e}$ averaged across random initial contingencies and random load scaling values.
As seen in the plot, the failure step error rate is in the order of 0.01 time steps. The significantly low error performance when averaged over random scaling values demonstrates the GNN's generalization capability.

\begin{figure}[htbp]
	\centering
	\begin{subfigure}[b]{\columnwidth}
		\centering
		\resizebox {\columnwidth} {!} {
\begin{tikzpicture}

\definecolor{darkblue00167}{RGB}{0,0,167}
\definecolor{darkgray176}{RGB}{176,176,176}
\definecolor{firebrick1933945}{RGB}{193,39,45}
\definecolor{lightgray204}{RGB}{204,204,204}

\begin{groupplot}[group style={group size=2 by 1}]
\nextgroupplot[
legend cell align={left},
legend style={fill opacity=0.8, draw opacity=1, text opacity=1, draw=lightgray204},
tick align=outside,
tick pos=left,
x grid style={darkgray176},
xlabel={Branch failure frequency},
xmin=-0.05, xmax=1.05,
xtick style={color=black},
y grid style={darkgray176},
ylabel={Failure time step error (generations)},
ymin=-0.0386315065556424, ymax=1.08506712323785,
ytick style={color=black}
]
\addplot [semithick, darkblue00167, dashed, mark=square*, mark size=2.25, mark options={solid}]
table {%
0 0.0812230442665225
0.1 0.0640026320381698
0.2 0.0501427656326664
0.3 0.0273406430575376
0.4 0.03430584638076
0.5 0.0519420896403869
0.6 0.045360979199648
0.7 0.0477511945390779
0.8 0.0598642726390515
0.9 0.0538883054174556
1 0.0124457038895163
};
\addlegendentry{Generalized GNN}
\addplot [semithick, firebrick1933945, dashed, mark=*, mark size=2.25, mark options={solid}]
table {%
0 0.632575757575758
0.1 1.03398991279269
0.2 0.74989837398374
0.3 0.544231834314633
0.4 0.364188450125755
0.5 0.332900159137059
0.6 0.257607871083291
0.7 0.202680248407385
0.8 0.232211171890235
0.9 0.175352076816914
1 0.0607162846763806
};
\addlegendentry{Influence model (1.40)}

\nextgroupplot[
legend cell align={left},
legend style={fill opacity=0.8, draw opacity=1, text opacity=1, draw=lightgray204},
tick align=outside,
tick pos=left,
unbounded coords=jump,
x grid style={darkgray176},
xlabel={Branch failure frequency},
xmin=-0.05, xmax=1.05,
xtick style={color=black},
y grid style={darkgray176},
ymin=-0.0278491111906466, ymax=0.620373767199555,
ytick style={color=black}
]
\addplot [semithick, darkblue00167, dashed, mark=square*, mark size=2.25, mark options={solid}]
table {%
0 0.0512254901960784
0.1 0.00161556509981707
0.2 0.0169358135989329
0.3 0.0247251234781514
0.5 0.00374799725337606
0.7 0.0151745068285281
0.8 nan
0.9 0.0027274280388359
1 0.00163019889190217
};
\addlegendentry{Generalized GNN}
\addplot [semithick, firebrick1933945, dashed, mark=*, mark size=2.25, mark options={solid}]
table {%
0 0.590909090909091
0.1 0.112919145160825
0.2 0.359559840850381
0.3 0.402552070350983
0.5 0.0481210556511761
0.7 0.091652030465185
0.8 nan
0.9 0.015065410421492
1 0.013924037556326
};
\addlegendentry{Influence model (1.40)}
\end{groupplot}

\end{tikzpicture} }
		\caption{Load scaling = $1.40$ for IEEE89 (left) and IEEE118 (right).}
	\end{subfigure}\vspace{4pt}
	\begin{subfigure}[b]{\columnwidth}
		\centering
		\resizebox {\columnwidth} {!} {
\begin{tikzpicture}

\definecolor{darkblue00167}{RGB}{0,0,167}
\definecolor{darkgray176}{RGB}{176,176,176}
\definecolor{firebrick1933945}{RGB}{193,39,45}
\definecolor{lightgray204}{RGB}{204,204,204}

\begin{groupplot}[group style={group size=2 by 1}]
\nextgroupplot[
legend cell align={left},
legend style={fill opacity=0.8, draw opacity=1, text opacity=1, draw=lightgray204},
tick align=outside,
tick pos=left,
x grid style={darkgray176},
xlabel={Branch failure frequency},
xmin=-0.05, xmax=1.05,
xtick style={color=black},
y grid style={darkgray176},
ylabel={Failure time step error (generations)},
ymin=-0.0340737203972498, ymax=0.990057932263815,
ytick style={color=black}
]
\addplot [semithick, darkblue00167, dashed, mark=square*, mark size=2.25, mark options={solid}]
table {%
0 0.0124777183600713
0.1 0.0442248479157656
0.2 0.023391874180865
0.3 0.0627693450929739
0.4 0.0513636596980744
0.5 0.0561465667457318
0.6 0.0351584397805616
0.7 0.0587555474152689
0.8 0.0757009406685867
0.9 0.0747692553684886
1 0.0214029460031817
};
\addlegendentry{Generalized GNN}
\addplot [semithick, firebrick1933945, dashed, mark=*, mark size=2.25, mark options={solid}]
table {%
0 0.943506493506494
0.1 0.66485657449889
0.2 0.546911179946836
0.3 0.472001817681898
0.4 0.368658096927846
0.5 0.307671162830917
0.6 0.247628317596455
0.7 0.311174427411759
0.8 0.226918608178947
0.9 0.19599573559415
1 0.0928065868885578
};
\addlegendentry{Influence model (1.50)}

\nextgroupplot[
legend cell align={left},
legend style={fill opacity=0.8, draw opacity=1, text opacity=1, draw=lightgray204},
tick align=outside,
tick pos=left,
unbounded coords=jump,
x grid style={darkgray176},
xlabel={Branch failure frequency},
xmin=-0.05, xmax=1.05,
xtick style={color=black},
y grid style={darkgray176},
ymin=-0.0200192605933791, ymax=0.455581933203257,
ytick style={color=black}
]
\addplot [semithick, darkblue00167, dashed, mark=square*, mark size=2.25, mark options={solid}]
table {%
0 0.00492610837438424
0.1 0.00191387559808612
0.2 0.0083827805162017
0.3 0.0372483084570041
0.4 nan
0.5 0.0223004894537741
0.6 0.00351091670864613
0.7 nan
0.8 0.00356385490313573
0.9 0.0015989754882862
1 0.00348451440612388
};
\addlegendentry{Generalized GNN}
\addplot [semithick, firebrick1933945, dashed, mark=*, mark size=2.25, mark options={solid}]
table {%
0 0.433963697121592
0.1 0.36026936026936
0.2 0.13900829197887
0.3 0.277381212058472
0.4 nan
0.5 0.128548233227826
0.6 0.0103078051202698
0.7 nan
0.8 0.0233360996685713
0.9 0.0110742096861961
1 0.0180099896242027
};
\addlegendentry{Influence model (1.50)}
\end{groupplot}

\end{tikzpicture} }
		\caption{Load scaling = $1.50$ for IEEE89 (left) and IEEE118 (right).}
	\end{subfigure}\vspace{4pt}
	\begin{subfigure}[b]{\columnwidth}
		\centering
		\resizebox {\columnwidth} {!} {
\begin{tikzpicture}

\definecolor{darkblue00167}{RGB}{0,0,167}
\definecolor{darkgray176}{RGB}{176,176,176}
\definecolor{firebrick1933945}{RGB}{193,39,45}
\definecolor{lightgray204}{RGB}{204,204,204}

\begin{groupplot}[group style={group size=2 by 1}]
\nextgroupplot[
legend cell align={left},
legend style={fill opacity=0.8, draw opacity=1, text opacity=1, draw=lightgray204},
tick align=outside,
tick pos=left,
unbounded coords=jump,
x grid style={darkgray176},
xlabel={Branch failure frequency},
xmin=-0.05, xmax=1.05,
xtick style={color=black},
y grid style={darkgray176},
ylabel={Failure time step error (generations)},
ymin=-0.0393264566853974, ymax=0.893088640534942,
ytick style={color=black}
]
\addplot [semithick, darkblue00167, dashed, mark=square*, mark size=2.25, mark options={solid}]
table {%
0 0.0115079365079365
0.1 0.0232512040546335
0.2 0.00826437328047515
0.3 0.0121924289455758
0.4 nan
0.5 nan
0.6 0.00738384169086352
0.7 0.0030560477337089
0.8 0.00639322287316667
0.9 0.0131473782249309
1 0.00334840269205165
};
\addlegendentry{Generalized GNN}
\addplot [semithick, firebrick1933945, dashed, mark=*, mark size=2.25, mark options={solid}]
table {%
0 0.0208333333333333
0.1 0.850706136115835
0.2 0.341963399648834
0.3 0.407562330551952
0.4 nan
0.5 nan
0.6 0.096841059602649
0.7 0.0463073633003959
0.8 0.0702274890787941
0.9 0.0557585005556605
1 0.0291805608313597
};
\addlegendentry{Influence model (1.90)}

\nextgroupplot[
legend cell align={left},
legend style={fill opacity=0.8, draw opacity=1, text opacity=1, draw=lightgray204},
tick align=outside,
tick pos=left,
unbounded coords=jump,
x grid style={darkgray176},
xlabel={Branch failure frequency},
xmin=-0.05, xmax=1.05,
xtick style={color=black},
y grid style={darkgray176},
ymin=-0.0131721383162449, ymax=0.296157508069132,
ytick style={color=black}
]
\addplot [semithick, darkblue00167, dashed, mark=square*, mark size=2.25, mark options={solid}]
table {%
0 0.04
0.1 0.00213219616204691
0.2 0.0156411788132218
0.3 0.0045519512801392
0.4 nan
0.5 0.00597701149425287
0.7 0.0249763523499852
0.8 0.000888300155817725
0.9 0.0024105352341442
1 0.00468964252371527
};
\addlegendentry{Generalized GNN}
\addplot [semithick, firebrick1933945, dashed, mark=*, mark size=2.25, mark options={solid}]
table {%
0 0.218253968253968
0.1 0.28209706959707
0.2 0.11992301738602
0.3 0.159049213008727
0.4 nan
0.5 0.0801469938461648
0.7 0.0911824625676139
0.8 0.0258734649746663
0.9 0.0223286970953464
1 0.0230145092311508
};
\addlegendentry{Influence model (1.90)}
\end{groupplot}

\end{tikzpicture} }
		\caption{Load scaling = $1.90$ for IEEE89 (left) and IEEE118 (right).}
	\end{subfigure}
	\caption{Branch failure steps prediction error rate $l_{failure{-}step,e}$ for IEEE89 (left) and IEEE118 (right) for various load scaling plotted against branch failure frequencies $l_{freq,e}$.}
	\label{fig:branchFailureSteps}\vspace{-8pt}
\end{figure}
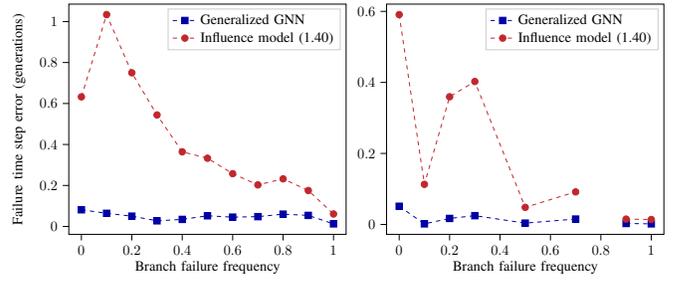
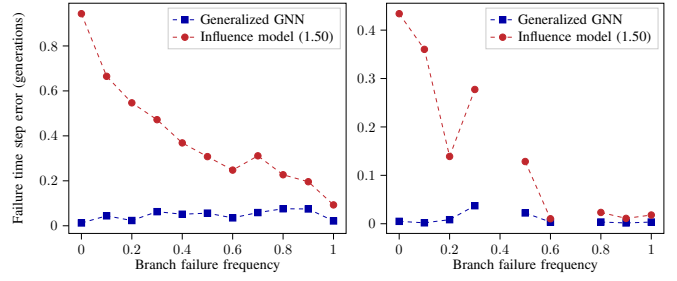
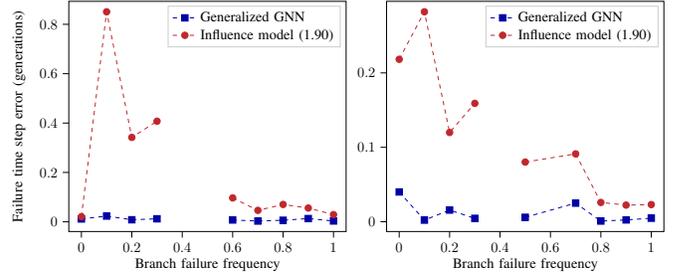

The average plot in Fig. \ref{fig:branchMetricFailureSteprandom} does not contain the performance of a reference influence model because of its inability to generalize over variable load profiles. Hence, to benchmark the performance of our model, Fig. \ref{fig:branchFailureSteps} plots the branch failure step error rates of the same GNN model, tested on three example load scaling values of $1.40$, $1.50$, and $1.90$, and compares it to different influence model instances built specifically for those scaling values of $1.40$, $1.50$, and $1.90$. 

As seen from the plots, the branch failure step prediction performance of the generic GNN model is significantly better than the load-specific influence model instances. In the influence model, when doing state prediction in a step-by-step manner, the errors that occur in initial steps propagate to later steps, thereby accumulating to a large final error. This is completely avoided by the GNN model since it predicts the failure steps directly. We believe this causes the GNN to outperform the influence models in this metric.


\subsubsection{Runtime Analysis}
We perform a runtime analysis to demonstrate how the GNN model can harvest the power of GPUs to predict cascade sequences much faster than the flow-based simulation methods. We run cascade predictions on 11,000 test samples with the CFS oracle, the influence model, and the GNN model. Table \ref{tab:timing_analysis} summarizes the runtime results. The CFS oracle cannot be run on a GPU, hence it was tested in MATLAB 2019a on an Intel(R) Core(TM) i9-7920X CPU@2.90GHz processor with 128GB of installed memory. Further, influence and GNN models were tested on an NVIDIA GeForce RTX 2080 Ti GPU with 11GB of total memory.\vspace{-6pt}
\begin{table}[htbp]
	\caption{Prediction time in seconds per 1000 samples.}
	\begin{center}
	\begin{tabular}{ c | c | c | c } 
		& CFS oracle & Influence model & GNN model\\ 
		\hline
		 IEEE89 & 24.18 & 2.35 & 0.53 \\ 
		IEEE118 & 62.54 & 1.86 & 0.28 \\ 
	\end{tabular}
	\label{tab:timing_analysis}
\end{center}
\end{table}\vspace{-12pt}

It can be seen that the time taken by the influence and GNN models are significantly lower than the CFS oracle. In the influence model, the matrix multiplications can be sped up using a GPU. However, because of its step-by-step prediction nature, each cascade prediction lasts for a variable number of steps. Hence, we cannot run multiple predictions simultaneously with the influence model unlike the fully parallel GNN model. Hence, the GNN model is almost four times faster than the influence model.



\section{Conclusion}
We considered the problem of predicting the failure cascade sequence due to branch failures given the initial contingency, power injection values, and grid topology. We proposed a flow-free graph neural network model that predicts the gird states at every generation of a cascade, without requiring power flow calculations. We showed that the model, in addition to being generic over randomly scaled loading values, outperforms the influence models that were built specifically for their corresponding loading profiles. Finally, we presented a runtime analysis to show that the model is faster by almost two orders of magnitude than the flow-based cascading failure simulator.

 \section*{Acknowledgment} This work was supported by NSF grants CNS-1735463 and CNS-2106268, and by a research award from the C3.ai Digital Transformation Institute.

\end{document}